\documentclass{article}

\usepackage{arxiv}

\usepackage{amsmath}
\usepackage{amssymb}
\usepackage{amsthm}
\usepackage{subcaption}

\usepackage[utf8]{inputenc} 
\usepackage[T1]{fontenc}    
\usepackage{hyperref}       
\usepackage{url}            
\usepackage{booktabs}       
\usepackage{amsfonts}       
\usepackage{nicefrac}       
\usepackage{microtype}      
\usepackage{cleveref}       
\usepackage{lipsum}         
\usepackage{graphicx}
\usepackage{natbib}
\usepackage{doi}

\def\Ttheta{\boldsymbol\theta}

\def\TC{\mathbf C}

\def\TK{\mathbf K}

\def\TY{\mathbf Y}

\def\TW{\mathbf W}

\def\TX{\mathbf X}

\newcommand{\bfA}{{\bf A}}
\newcommand{\bfB}{{\bf B}}

\newcommand{\bfK}{{\bf K}}

\newcommand{\bfX}{{\bf X}}
\newcommand{\bfY}{{\bf Y}}

\newcommand{\bfx}{{\bf x}}
\newcommand{\bfy}{{\bf y}}
\newcommand{\bfu}{{\bf u}}

\newcommand{\bfv}{{\bf v}}

\newcommand{\bftheta}{{\boldsymbol \theta}}

\newtheorem{theorem}{Theorem}[section]

\title{Fully invertible hyperbolic neural
networks for segmenting large-scale surface and sub-surface
data}

\newif\ifuniqueAffiliation
\uniqueAffiliationtrue

\ifuniqueAffiliation 
\author{ Bas Peters\\
	Computational Geosciences Inc.\\
	Vancouver, BC\\
	\texttt{bas@compgeoinc.com} \\
	\And
	Eldad Haber\\
	Department of Earth, Ocean, and Atmospheric Sciences\\
	The University of British Columbia\\
	Vancouver, BC\\
	\AND
 	Keegan Lensink\\
	Department of Earth, Ocean, and Atmospheric Sciences\\
	The University of British Columbia\\
	Vancouver, BC\\
}
\else
\fi

\renewcommand{\shorttitle}{Fully invertible hyperbolic neural
networks}

\hypersetup{
pdftitle={FHNN},
pdfsubject={physics.geo-ph},
pdfauthor={Bas Peters},
pdfkeywords={Invertible Neural Networks, Large Scale Deep Learning, Memory Efficient Deep Learning},
}

\begin{document}
\maketitle

\begin{abstract}
The large spatial/temporal/frequency scale of geoscience and remote-sensing datasets causes memory issues when using convolutional neural networks for (sub-) surface data segmentation. Recently developed fully reversible or fully invertible networks can mostly avoid memory limitations by recomputing the states during the backward pass through the network. This results in a low and fixed memory requirement for storing network states, as opposed to the typical linear memory growth with network depth.
This work focuses on a fully invertible network based on the telegraph equation. While reversibility saves the major amount of memory used in deep networks by the data, the convolutional kernels can take up most memory if fully invertible networks contain multiple invertible pooling/coarsening layers. We address the explosion of the number of convolutional kernels by combining fully invertible networks with layers that contain the convolutional kernels in a compressed form directly. A second challenge is that invertible networks output a tensor the same size as its input. This property prevents the straightforward application of invertible networks to applications that map between different input-output dimensions, need to map to outputs with more channels than present in the input data, or desire outputs that decrease/increase the resolution compared to the input data. However, we show that by employing invertible networks in a non-standard fashion, we can still use them for these tasks. Examples in hyperspectral land-use classification, airborne geophysical surveying, and seismic imaging illustrate that we can input large data volumes in one chunk and do not need to work on small patches, use dimensionality reduction, or employ methods that classify a patch to a single central pixel.
\end{abstract}


\section{Introduction}
Many datasets in the imaging sciences are intrinsically 3D or 4D. For instance, the interpretation of seismic imagery, hyperspectral data segmentation for land-use classification, and segmentation of various medical imagery. To construct convolutional neural networks with a sufficiently large field-of-view (receptive field), we need deeper networks with more layers or multiple coarsening (pooling) and refinement stages in the network. Reasons we wish to work on large chunks of data instead of many small patches include wanting to learn from larger length scales that may be present in the data, as well as weakly supervised approaches that add prior knowledge or constraints related to properties of full images \citep{KERVADEC201988,peters2022point,doi:10.1142/S0219530519410148}.

The dominant factor that limits the input data size and network depth is the storage of the network state, that is, the convolved data at each layer that is needed in order to compute a gradient of the loss function using back-propagation, implemented via reverse-mode automatic differentiation. Re-computing the network (forward) states in reverse order during back-propagation avoids this problem. This re-computation is possible when using fully invertible (also known as reversible) networks. Fully invertible networks have a constant memory requirement for states (activations) that is independent of network depth and the number of pooling stages, see Figure \ref{fig:memorycurves}. Therefore, fully invertible networks largely avoid the memory limitations related to storing states. 

Specifically, we employ a second-order hyperbolic differential equation based invertible network \citep{lensink2019fully}. The connection with the wave equation enables the use of a suite
of tools from analysis to interpretations that are commonly used in mathematical physics and numerical analysis. We note that other invertible network constructions exist, including invertible Hamiltonians \citep{RuthottoHaber2018}, invertible ResNets \citep{pmlr-v97-behrmann19a}, invertible u-nets \citep{9231874}. The fully invertible networks generally extend invertible networks for image classification \citep{jacobsen2018irevnet,leemput2018memcnn} and networks that are only invertible in between coarsening/pooling stages \citep{RuthottoHaber2018, Chang2017Reversible,GomezEtAl2017,DinhSB16}. 

Fully invertible networks use invertible pooling/coarsening operations. Examples are the Haar transform \citep{lensink2019fully}, reordering via a checkerboard pattern \citep{DinhSB16,jacobsen2018irevnet}, or various learned coarsening operators \citep{lensink2019fully,9231874}. The invertible pooling causes the fully hyperbolic invertible network to be less flexible than some other networks for image segmentation in two ways. First, the convolutional kernels become the dominant memory consumer when the network contains several down-sampling operators. A fully invertible hyperbolic network needs to increase the number of channels by a factor of eight to change the resolution by a factor of two in each direction in 3D. This preservation of the number of elements in the tensors makes the coarsening and channel-count changes invertible operations. However, this approach leads to an `explosion' of the channels, as remarked by \citet{peters2019symmetric, 9231874}. For instance, if the input is three-channel RGB, there are $192$ channels after two coarsening layers and an astonishing $98304$ channels in case we wish to coarsen five times. The storage and computations of the associated $98304^2$ convolutional kernels (just for one layer at the coarsest level) would be completely unfeasible. Figure \ref{fig:memorycurves} illustrates this effect.

A second way in which the fully invertible hyperbolic network is a relatively `rigid' design is that an orthogonal transform can increase and later decrease the number of channels by $8\times$ per coarsening layer only, but we cannot arbitrarily reduce the number of channels and, thus network parameters. Similarly, one cannot increase the number of channels or make the network wider without decreasing resolution. This will often cause the network to contain too many or too few convolutional kernels for a certain task, given the network depth and the number of coarsening/refinement stages. The standard design of an invertible network outputs a tensor of the same size as its input and with the same number of channels. Therefore, one cannot directly apply invertible networks to applications like hyperspectral imaging that map 3D/4D inputs to a 2D output. Other applications that cannot directly work with invertible networks include ones that need to map to outputs with more channels than present in the input data or desire outputs that decrease/increase the resolution compared to the input data.

In this work, we present solutions to the above problems by combining the design of fully invertible networks with layers that can reduce the storage and computations related to the convolutional kernels. The same layer also serves as a way to increase the number of convolutional kernels per layer if required, without changing resolution. These two features make fully invertible hyperbolic networks much more flexible and fix the primary disadvantages. Furthermore, we show that we can, in fact, use invertible neural networks to change the resolution or the number of output channels while maintaining network invertibility.

Several examples illustrate how the presented tools enable the application of invertible neural networks to the following geoscientific problems: 1) time-lapse hyperspectral land-use-change detection, which maps 4D data to a 2D map; 2) large-scale 2D multi-model airborne-geophysical and remote sensing for aquifer mapping, where we map from dozens of input channels to a couple of output classes; 3) geological model building from seismic data, where we set up the network so that it outputs a lower resolution compared to the input.

The examples illustrate that the developed tools extend the type of problems that can be handled using fully hyperbolic architectures, enabling training using larger data blocks as input for the network. This, in turn, allows us to work with higher-resolution inputs and learn larger-scale (spatial/temporal/harmonic) patterns that are present in the data.

\subsection{Contributions}

This work looks at some practical obstacles when applying fully invertible neural networks based on hyperbolic PDEs to large-scale remote sensing and geoscience problems. Specifically, we note our primary contributions as:
\begin{itemize}
\item To the best of our knowledge, this is the first work that addresses the issue of the `exploding' memory for convolutional kernels with an increasing number of resolution/pooling changes in a hyperbolic invertible network. Our solution keeps the network fully invertible while drastically reducing the number of convolutional kernels and associated memory and thus enables learning from data on a much larger scale than before while also working with arbitrarily deep networks.

\item We present a few subtle modifications that remove the limitation that fully invertible hyperbolic networks map between inputs and outputs of the same size/resolution and channel count while not giving up full invertibility and without increasing the computational cost or memory requirements.
\end{itemize}

After reviewing the design and some properties of the invertible hyperbolic neural network, we illustrate the limitations of the network structure. Then, we propose our solutions. Finally, we train on hyperspectral, multi-modality, and seismic datasets with sparse spatial label sampling.

\section{Fully invertible hyperbolic neural networks for large input-output problems}

The overwhelming majority of the literature relies on reverse-mode automatic differentiation for gradient computation. This type of gradient computation requires access to the network states $\TY_j$ at layer $j$ during the backpropagation phase. Standard implementations keep all network states in memory, causing the memory footprint to grow linearly with network depth. Workarounds to reduce memory may rely on shallow networks or a network design that maps a small patch or data sub-volume into the class of the central pixel/voxel. 

Fully invertible networks based on PDEs require memory for just a couple of layer states, depending on the discretization. So, there is no longer a need to trade off network depth for input size. Memory savings by using invertible architectures allow us to allocate all available memory towards larger data input volumes, enabling the network to learn from large-scale structures in the data.

As discussed in the introduction, out of the various fully invertible network designs, our focus is on the physics-inspired invertible network based on the non-linear Telegraph equation \citep{Zhou2018Telegraph} with time-step $h$. This equation is the basis for the invertible architecture of \citet{RuthottoHaber2018,Chang2017Reversible}.   
 \begin{eqnarray}
 \label{telegraph}
\frac{\partial^2 \bfY}{\partial t^2} = f(\bfY,\bftheta(t)),
 \end{eqnarray}
where $\bfY$ is the state, $f$ is a non-linear function, and the model parameters $\bftheta(t)$ are time dependent. The model parameters often parameterize convolutional, block-convolutional, differential and dense matrices, denoted as $\bfK(\bftheta(t))$. While many neural networks use nonlinearities of the type $ f(\bfY(t),\bfK(\bftheta(t))) = \sigma (\bfY(t),\bfK(\bftheta(t)))$ with a nonlinear, point-wise, and monotonically increasing activation function $\sigma : \mathbb{R} \rightarrow \mathbb{R}$, we select the symmetric layer \citep{RuthottoHaber2018} of the form
\begin{equation}\label{symmetric_layer}
f(\bfY,\bftheta(t)) = -\bfK(\bftheta(t))^{\top} \sigma(\bfK(\bftheta(t)) \bfY(t)).
\end{equation}
With this choice, \eqref{telegraph} becomes
\begin{equation}\label{telegraph2}
\frac{\partial^2 \bfY}{\partial t^2} = -\bfK(\bftheta(t))^{\top} \sigma(\bfK(\bftheta(t)) \bfY(t)).
\end{equation}
The motivation for the specific symmetric layer choice relates to stability and energy conservation of the forward propagation through the network; see \citet{RuthottoHaber2018} for a stability proof.

To proceed, we follow \citet{RuthottoHaber2018} and use the conservative Leapfrog discretization of the second derivative of the state, 
\begin{equation}\label{LeapFrog}
\frac{\partial^2 \bfY}{\partial t^2}  \approx {\frac 1 {h^2}} \left(\bfY_{j+1} - 2 \bfY_j + \bfY_{j-1} \right),
 \end{equation}
where $h$ now indicates the artificial-time step and $j$ indexes the discrete time. Combining the discretization of the second derivative and \eqref{telegraph2} results in
\begin{align}\label{network0}
\TY_1 = \: &\TX, \quad \TY_2 = \TX \nonumber \\
\TY_{j} = \: &2 \TY_{j-1} -   \TY_{j-2} -  
h^2 \TK(\Ttheta_{j})^\top \sigma( \TK(\Ttheta_{j})  \TY_{j-1}) 
\end{align}
Equation~\eqref{network0} is a hyperbolic network that uses
a single resolution. 
The first two states are the initial conditions, which we set equal to the input data $\TX \in \mathbb{R}^{n_1 \times n_2 \times n_3 \times n_\text{chan}}$. The data tensor has $n_\text{chan}$ channels and 3 other dimensions indexed by $n_1$, $n_2$, and $n_3$. Examples include hyperspectral data and 3D seismic image volumes.   

An artificial time-step $h$ affects the stability of the forward propagation \citep{HaberRuthotto2017a} via the well-known CFL condition \citep{leveque}. In this work, we set the linear operator $\TK(\Ttheta_j)$ to convolutions with kernels $\Ttheta_j$, and select the ReLU as the activation function $\sigma(\cdot)$ for the examples.

\begin{figure}
    \centering
    \includegraphics[width=0.75\linewidth]{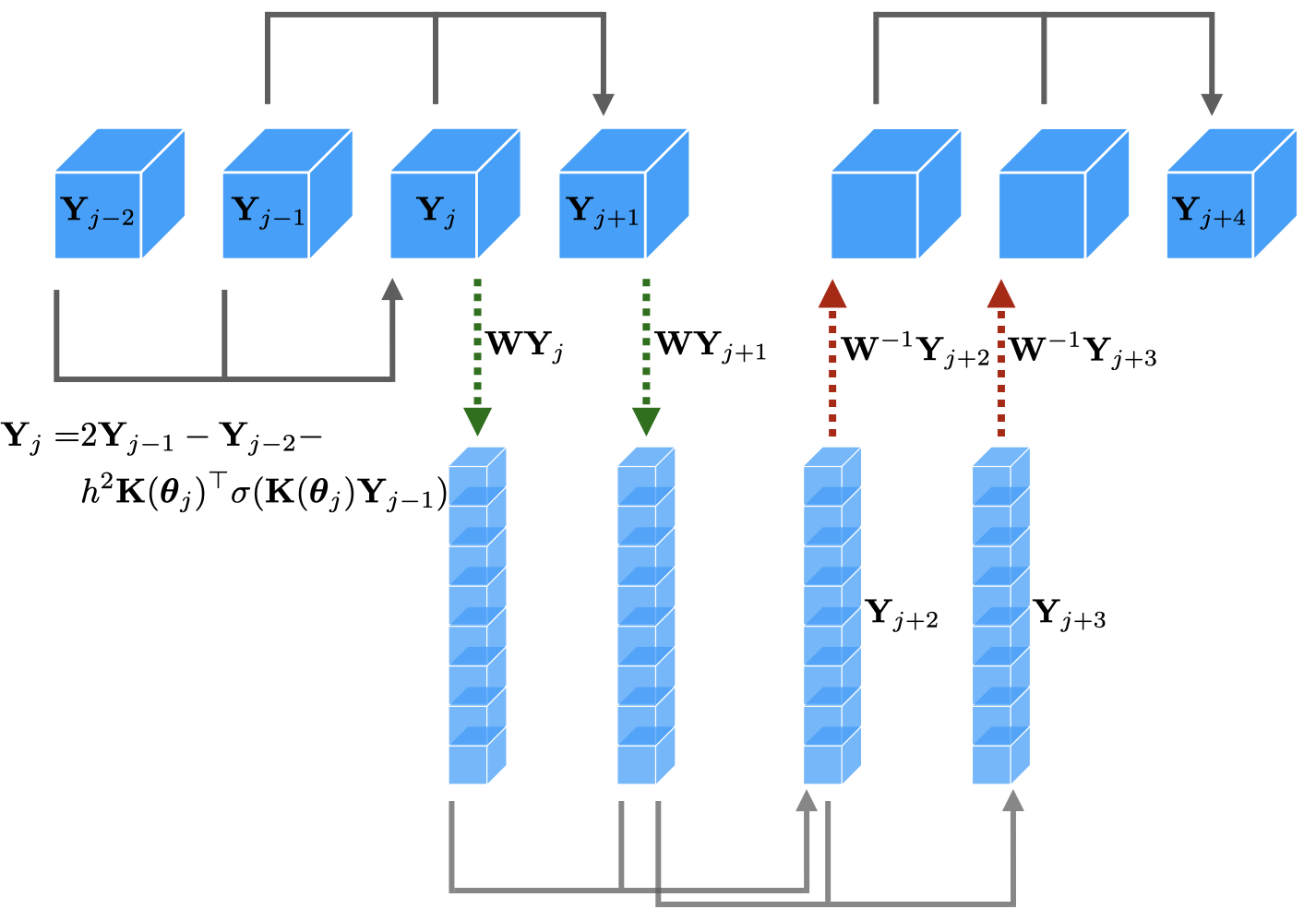}
    \caption{Diagram of the flow of the multi-level hyperbolic network in \eqref{network}. Shown for 3D input data with one channel and two levels. The network contains seven layers, with pooling after layer four and unpooling after layer six.}
    \label{fig:HyperNetIllustration}
\end{figure}

In order to introduce multi-resolution into the system
we use the approach by \citet{lensink2019fully} and introduce the
linear operators $\TW_j$ that change the resolution of the
state {\bf without} losing information by moving the information from the spatial dimension to the channel dimension, obtaining the network 
\begin{align}\label{network}
\TY_1 = \: &\TX, \quad \TY_2 = \TX \nonumber \\
\TY_{j} = \: &2 \TW_{j-1}\TY_{j-1} -  \TW_{j-2} \TY_{j-2} - h^2 \TK(\Ttheta_{j})^\top \sigma( \TK(\Ttheta_{j}) \TW_{j-1} \TY_{j-1}), \: j=3,\cdots,n.
\end{align}
 The operator $\TW$ represents coarsening/pooling to change the resolution and the number of channels simultaneously. Note that $\TW_j$ equals the identity if there is no resolution change at layer $j$. Figure \ref{fig:HyperNetIllustration} illustrates an instance of the network design.

The operators $\TW_j$ are important components of networks for image-to-image mappings, and they need to be invertible operators to enable full invertibility of the network. Practical invertible linear operators are orthogonal, do not require storage of their dense matrix representation, and have fast forward and inverse transforms known in closed form. Here, we select the orthogonal Haar wavelet transform \citep{WaveletReview} $\TW$ to coarsen the image and increase the number of channels simultaneously. This choice was used successfully in \citet{lensink2019fully}. The transpose (and inverse) achieves the reverse of these operations, $\TW^\top=\TW^{-1}$ and $\TW^T \TW =I = \TW \TW^\top$, so the action of the linear operator $\TW$ on a tensor $\TY$ creates the mappings
\begin{align}
\TW \TY &: \mathbb{R}^{n_1 \times n_2 \times n_3 \times n_\text{chan}} \rightarrow \mathbb{R}^{n_1/2 \times n_2/2 \times n_3/2 \times 8n_\text{chan}}, \label{eq:Wtransform}\\
\TW^{-1} \TY&: \mathbb{R}^{n_1 \times n_2 \times n_3 \times n_\text{chan}} \rightarrow \mathbb{R}^{2 n_1 \times 2 n_2 \times 2 n_3 \times n_\text{chan}/8}.\label{eq:iWtransform}
\end{align}
Because of the invertibility of any orthogonal transform, applying $\TW$ and $\TW^\top$ incurs no loss of information. As mentioned in the introduction, other invertible pooling or coarsening/refining operators are also available.

Invertibility of the full network is exploited by isolating one of the states in \eqref{network} and reversing the indices to obtain an expression for the current state in terms of future states:
\begin{align}\label{rev_prop}
\TY_j = \TW_j^{-1} \bigg[ &2 \TW_{j+1} \TY_{j+1} - h^2 \TK(\Ttheta_{j+2})^\top  \sigma ( \TK(\Ttheta_{j+2}) \TW_{j+1} \TY_{j+1} ) - \TY_{j+2} \bigg].
\end{align}
This equation does not require inverting the activation function $\sigma$. Instead, only the inversion of the orthogonal wavelet transform is required, which is known in closed form. When computing the gradient of the loss function using backpropagation, we recompute the states $\TY_j$ while propagating backwards through the network. The re-computation avoids the storage of all $\TY_j$ and leads to a fixed memory requirement for the states of three layers, see Table \ref{memory} for an overview.

\begin{table}[]
\caption{Memory requirements for the states $\TY_j$ and convolutional kernels $\Ttheta$ for fully invertible and non-invertible equivalent networks based on the networks in Table \ref{network_design}.}
\label{memory}
\begin{center}
\begin{tabular}{lcl}
\multicolumn{1}{c}{Memory}  &\multicolumn{1}{c}{\bf States} &\multicolumn{1}{l}{\bf Conv. kernels}
\\  \hline \\ \hline \\
\bf Hyperspectral \\
non invertible          & $22.5$ GB& $0.02$ GB\\
invertible              & $3.7$ GB& $0.02$ GB\\
invertible + BLR layers & $3.7$ GB& $0.003$ GB\\
\hline \\
\bf Aquifer mapping \\
non invertible                & $21.02$ GB& $32.19$ GB\\
invertible             & $1.66$ GB& $32.19$ GB\\
invertible + BLR layers & $1.66$ GB& $0.12$ GB\\
\hline \\
\bf 3D seismic \\
non invertible          & $21.96$ GB & $41.16$ GB\\
invertible              & $2.19$   GB & $41.16$ GB\\
invertible + BLR layers & $2.19$   GB & $0.23$ GB\\
\hline
\end{tabular}
\end{center}
\end{table}

\begin{figure}[!t]
\begin{center}
   \includegraphics[width=0.9\linewidth]
                   {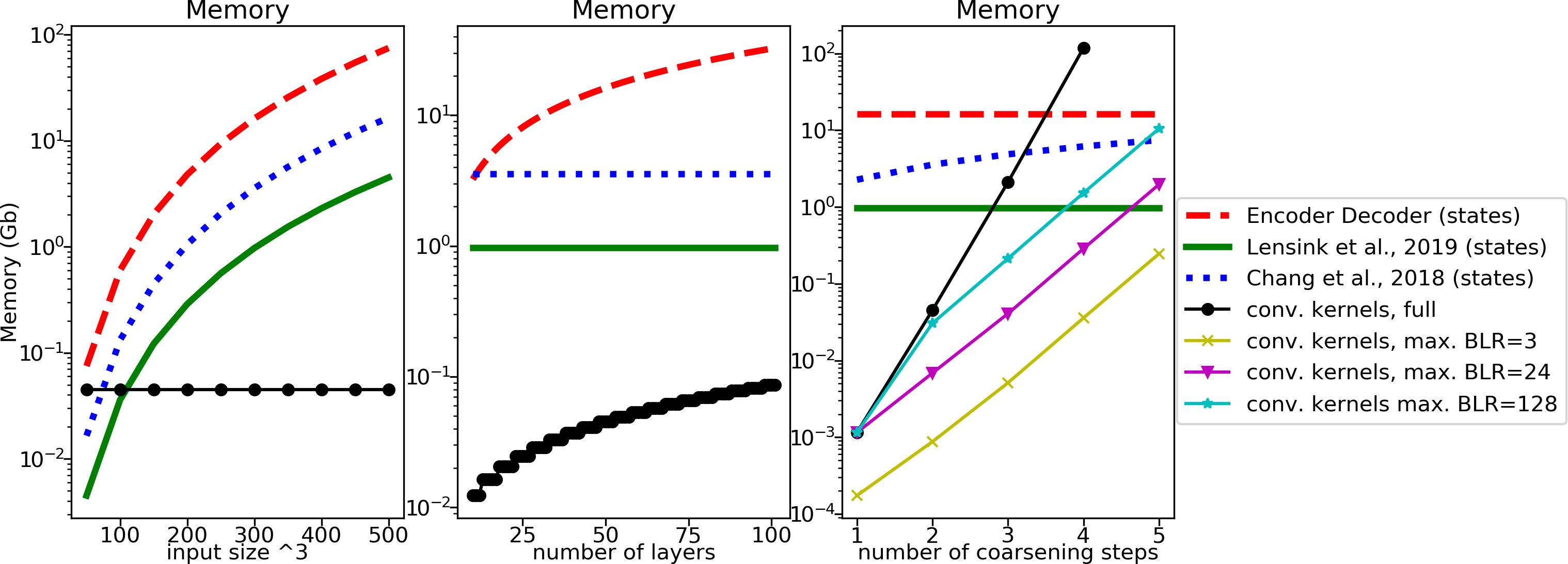}
\end{center}
   \caption{Memory requirements (Gigabyte) for network states (activations) and $3 \times 3 \times 3$ convolutional kernels. Left: as a function of input size and a fixed $50$ layer network with two coarsening stages. Middle: as a function of an increasing number of layers but with fixed input size ($300^3$) and fixed number of two coarsenings. Right: as a function of an increasing number of coarsening steps but with a fixed number of layers ($50$) and input size ($300^3$). Our proposed Block-Low-Rank (BLR) layers avoid an exploding number of convolutional kernels with increased coarsening in invertible networks.}
\label{fig:memorycurves}
\end{figure}

The benefits of the hyperbolic network design range beyond just invertibility and memory savings. We point out that the network design described in this section can include instance/batch/layer normalization. However, we found it unnecessary for our examples. The following stability property from \citep[Thm. 2]{RuthottoHaber2018} can partially explain this observation.

\begin{theorem}[Stability of the fully invertible hyperbolic network \eqref{telegraph2} and \eqref{network}.]\label{thm1}
{
The neural network satisfies the stability criterion 
\begin{equation}
\| \bfy_1(t=T) - \bfy_2(t=T) \|_2^2 \leq c \| \bfy_1(t=0) - \bfy_2(t=0) \|_2^2
\end{equation}
where $\bfy_1$ and $\bfy_2$ are two different initial states, given at times equal to zero and after propagating to time $T$. The constant $c>0$ is independent of $t$.
}
\end{theorem}
The above theorem is stated in \cite[Thm. 2]{RuthottoHaber2018} for time-independent weights $\TK$. Without stating the full proof, we illuminate the structure of the hyperbolic network, stability, and connection to other physical wave-equations. To start, rewrite \eqref{telegraph2} into a first-order system of equations by introducing the state $\bfv(t)$ and using $\bfK \bfu(t) = \partial_t \bfv(t)$ and $-\bfK^\top \bfv(t) = \partial_t \bfu(t)$ to arrive at
\begin{align}\label{eq:first_order_system}
\partial_t \begin{bmatrix} \bfu(t) \\ \bfv(t) \end{bmatrix}
&=
\begin{bmatrix} I & 0 \\ 0 & \sigma(\cdot) \end{bmatrix}  \begin{bmatrix}0 & -\bfK^\top \\ \bfK & 0 \end{bmatrix} \begin{bmatrix} \bfu(t) \\ \bfv(t) \end{bmatrix}\\
&\leftrightarrow \partial_t \bf
\bfx(t)= \bfB \circ \bfA \bfx(t).\nonumber
\end{align}
The above immediately shows that the initial choice of network non-linearity $ -\bfK(t)^{\top} \sigma(\bfK(t) \bfY(t))$ leads to the anti-hermitian operator, or skew-symmetric matrix property $A^\top = -A$. This property, in combination with \ref{eq:first_order_system}, leads to a standard stability argument via the upper bound of the energy
\begin{equation}\label{eq:Ebound}
\| \bfB \circ \bfA \bfx(t) \|_2^2 \leq \| \bfA \bfx(t) \|_2^2 \leq \| \bfA \|_2^2 \| \bfx(t) \|_2^2 = c \| \bfx(t) \|_2^2,
\end{equation}
where the nonlinearity satisfies $| \sigma(\bfx) | \leq | \bfx | $ and $c$ is the constant $\| \bfA \|_2^2$. This energy definition is equivalent to the one in \cite[sect. 2.4.3]{evans2010partial} and \cite{RuthottoHaber2018}. The upper bound is the energy associated with the linear wave-like equation. The energy of a linear wave-like equation is constant in time, as shown by a standard argument using the anti-hermitian $\bfA$ as follows:
\begin{align}\label{eq:energy_conservation}
\frac{\partial \| \bfx(t) \|_2^2}{\partial t} &= \frac{\partial }{\partial t}(\bfx(t),\bfx(t)) 
= (\partial_t \bfx(t), \bfx(t)) + (\bfx(t),\partial_t \bfx(t)) \nonumber \\
&= (\bfA \bfx(t),\bfx(t)) + (\bfx(t), \bfA \bfx(t)) \nonumber \\
&= (\bfA \bfx(t), \bfx(t)) - (\bfA \bfx(t), \bfx(t)) = 0
\end{align}
where the norm is induced by the inner product $\| \bfx \|_2^2 = (\bfx,\bfx)$.

Thus, the energy of the hyperbolic network, in between resolution changes, is bounded from above by the energy of the linear wave-like equation, which itself is bounded. The orthogonal wavelet transforms also do not change the energy across resolution levels.

Below, we will verify the stability property for randomly initialized and trained networks.

\subsection{Empirical verification of stability properties}
\begin{figure}[!t]
\begin{center}
   \includegraphics[width=0.9\linewidth]
                   {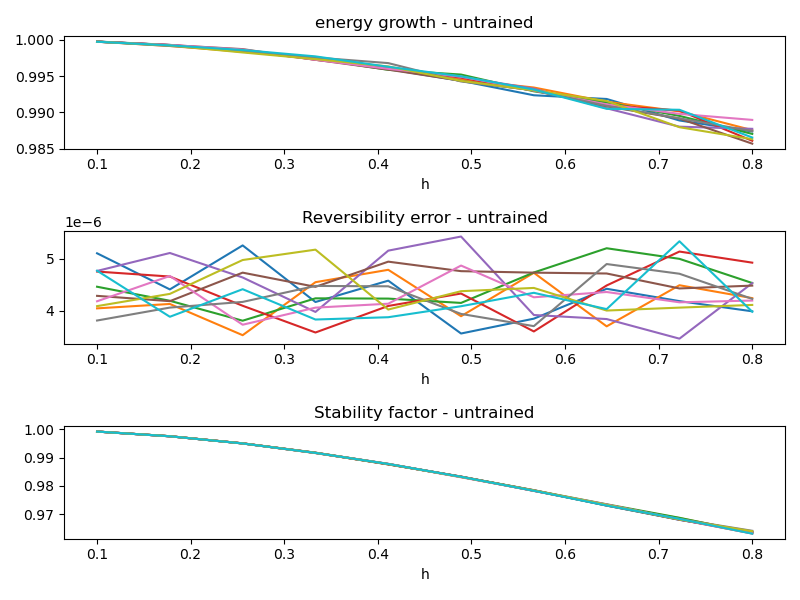}
\end{center}
   \caption{Empirically observed properties of an untrained and randomly initialized invertible hyperbolic network, as a function of the `time-step' $h$. The perturbation for the middle figure is also chosen randomly for every test point.}
\label{fig:empiricalstability}
\end{figure}

The inherited stability of the invertible hyperbolic network as stated above suggests it is possible to train the networks for the examples without using any instance/batch/layer normalization. Here, we illustrate the stability, as well as invertibility and energy growth, for both randomly initialized and trained networks. Figure \ref{fig:empiricalstability} displays empirically observed network properties for an untrained version of the ($30$ layer, four-level) network described in Table \ref{network_design_seismic}, for a range of artificial time-step sizes $h$. The function $g(\bfX,\bftheta)$ denotes the propagation of input $\bfX$ through the network, and $g^{-1}$ denotes the inverse/reverse propagation. The figures show the following properties:\\

\begin{itemize}
\item Energy growth: $\frac{\| g(\bfX_0,\bftheta) \|_2}{\| \bfX_0 \|_2}$, which is the ratio of the norms of network input and output.

\item Stability: $\frac{\| g(\bfX,\bftheta) - g(\bfX+\delta \bfX,\bftheta) \|_2 }{\| \delta \bfX \|_2}$, where $\delta \bfX$ is a random perturbation (selected with a magnitude of $0.1 \| \bfX \|_2$ for the current experiment).

\item Invertibility error: $\frac{\| g^{-1}(g(\bfX,\bftheta), \bftheta ) - \bfX \|_2}{\| \bfX \|_2}$
\end{itemize}

The results show that a network initialized with zero mean and normally distributed weights is stable, preserves energy relatively well, and is invertible in practice. For energy growth and stability, a U-net with the same number of levels, layers per level, and the same number of channels resulted in exploding outputs. We note that the results depend on the scaling of the random initialization and that a smart initialization like \cite{pmlr-v9-glorot10a} can also improve the stability at the start of training. It can be verified that our network shows similarly good results for a wide range of scalings of random initial weights. 

The same test after training the network revealed the following numbers for each property: energy growth: $7.42$, which is a reasonably small number (note that energy growth is not expected to be smaller than one, see \eqref{eq:Ebound}); stability: mean of $1.05$ with std $0.004$; invertibility error: $0.00018$. These numbers illustrate the favorable stability, energy growth and invertibility properties are preserved while training the network.

\section{Problems and solutions for practical invertible networks}
In the following, we propose solutions to the limitations of fully invertible multi-level hyperbolic networks as posed in the introduction: an exploding number of convolutional kernels, inability to deal with different input-output resolutions and channel numbers, and transformations between different data-label dimensions. Below, we address each of these problems and present solutions such that we can apply hyperbolic invertible networks anyway, without fundamentally altering the network design or giving up invertibility.

\subsection{Exponential parameter growth with the number of resolution changes - a block low-rank approach}

Orthogonal transforms that make neural networks fully invertible also preserve the number of entries after pooling/coarsening and unpooling/refining the state. Indeed, equations \eqref{eq:Wtransform} and \eqref{eq:iWtransform} show that if there are $n_\text{chan}^2$ convolutional kernels per network layer, starting with 3D input data with $n_\text{chan (ini)}$ channels leads to

\begin{equation}
n_\text{chan} = n_\text{chan (ini)} \times 8^{n_\text{coarsening}}
\end{equation}

channels after coarsening/pooling $n_\text{coarsening}$ times. Equivalently, each time the network states coarsen, the number of convolutional kernels grows with a factor of $64$. See Figure \ref{fig:memorycurves} for an illustration of this effect for various network designs as a function of input size, number of layers, and the number of coarsening steps. As noted by \cite{peters2019symmetric, 9231874}, this results in prohibitive memory and computation times for storing and computing convolutional kernels, respectively.

From the above observations and Figure \ref{fig:memorycurves}, it is clear that invertible hyperbolic networks with multiple coarsening stages can only be a practical tool if it is possible to reduce the memory requirements for the network parameters. Various memory reduction techniques for network weights have been developed, and below we discuss a few that are/are not suitable for our purposes. We also introduce a novel network layer defined by a low block rank.

To start, consider network compression methods such as pruning and low-rank matrix/tensor factorization \citep{NIPS2014_5544} train a 'full' network first, followed by compression of the weights. These ideas do not apply to our situation because we assume that it is infeasible to train the full network. Instead, we need methods that directly train a reduced-memory network. Prior work includes replacing a part of the convolutional kernels by scalars \citep{ephrath2019leanresnet}, and equipping the convolutional kernels with a block-circulant structure for parameter reduction \citep{8686678,treister2018low}, as well as employing Kronecker-product structures \citep{kron_approx_network}. 

Here, we introduce another layer type that requires minimal adaptation to existing code, has easily adaptable memory requirements, and has a clear linear-algebraic interpretation. Inspired by techniques like LR-factorization for matrix completion \citep{rennie2005fast,doi:10.1137/130919210}, we construct layers with a block-low-rank structure formed by the convolutional kernels when structured in its flattened matrix form. This structure allows for explicit limitations on the number of convolutions in a layer while preserving a possibly extremely large number of channels. Our approach is similar to `squeeze-and-expand' bottleneck methods like \citet{Szegedy_2015_CVPR,he2016deep,iandola2016squeezenet} proposed for non-invertible residual networks, mainly in the context of image classification. Our layer differs because we have one instead of two nonlinearities per layer, and our convolutional kernels have a symmetric structure (see Eq. \eqref{symmetric_layer}), which squeezenet does not have. If the nonlinearity is not symmetric, it will not induce an anti-hermitian operator in \ref{thm1} and, therefore, cannot guarantee stability. 

Also different is that we do not rely on $1 \times 1 \times 1$ convolutions for compression and expansion. Because we explicitly induce and recognize the block-low-rank structure in our network layer, we can directly observe various linear-algebraic properties of interest.

Linear algebraic matrix-vector product notation is required to interpret and construct the block-low-rank layer. The computational implementation can still use a 5D tensor format for the numerical examples. First, flatten the tensors that contain network states of type $\mathbb{R}^{n_x \times n_y \times n_z \times n_\text{chan} }$ to block-vectors $\mathbb{R}^{n_x n_y n_z n_\text{chan}}$ that contain $n_\text{chan}$ sub-vectors $Y^i \in \mathbb{R}^{n_x n_y n_z}$,
\begin{equation}
\TY \equiv
\begin{bmatrix}
Y^1 \\
Y^2 \\
\vdots \\
Y^{n_\text{chan}}
\end{bmatrix}.
\end{equation}
 
Similarly, we rewrite the convolutional kernels for a given layer in tensor format $\mathbb{R}^{n_x \times n_y \times n_z \times n_\text{chan out} \times n_\text{chan in}}$ as the block matrix $\TK$ with $n_\text{chan out}$ rows and $n_\text{chan in}$ columns. Each block $K(\theta^{i,k})$ is a Toeplitz matrix representation of the convolution with a kernel $\theta^{i,k}$, 
\begin{equation}
\TK \equiv 
\begin{bmatrix}
K(\theta^{1,1}) & K(\theta^{1,2}) & \hdots & K(\theta^{1,n_\text{chan in}}) \\ 
K(\theta^{2,1}) & K(\theta^{2,2}) & \hdots & K(\theta^{2,n_\text{chan in}}) \\ 
\vdots & \vdots & \ddots & \vdots \\
 K(\theta^{n_\text{chan out},1}) & K(\theta^{n_\text{chan out},2}) & \hdots & K(\theta^{n_\text{chan out},n_\text{chan in}}) \\ 
\end{bmatrix}.
\end{equation}
This block matrix is square if the number of channels remains unchanged after the convolutions. To reduce the number of channels, $\TK$ needs to be a flat matrix; a tall block-matrix increases the number of channels. The collection of convolutional kernels at layer $j$ is denoted by $\Ttheta_j$.

Using this notation, the non-linear part of each layer in the hyperbolic network \eqref{network} becomes (as an example, consider a case with three input channels in $\bfY_j$ but using only six convolutional kernels instead of the usual nine)

\begin{equation}\label{BLRLayer}
\begin{bmatrix}
K(\theta^{1,1})^\top & K(\theta^{2,1})^\top\\
K(\theta^{1,2})^\top & K(\theta^{2,2})^\top\\
K(\theta^{1,3})^\top & K(\theta^{2,3})^\top
\end{bmatrix}
\sigma \Bigg(
\begin{bmatrix}
K(\theta^{1,1}) & K(\theta^{1,2}) & K(\theta^{1,3})\\
K(\theta^{2,1}) & K(\theta^{2,2}) & K(\theta^{2,3})\\
\end{bmatrix}
\begin{bmatrix}
Y^1 \\
Y^2 \\
Y^3 \\
\end{bmatrix}
\Bigg).
\end{equation}

For simplicity, we assumed there is no coarsening via the Haar transform at this particular layer. The symmetric layer structure enables a layer to have the same number of input and output channels while working with fewer convolutional kernels than $n_\text{chan}^2$ if $\TK$ is a tall block-matrix.

A benefit of our proposed construction is the following clear linear-algebraic interpretation:

The structure $\bfK_j^{\top} \sigma (\bfK_j \bfY_j )$ using a block matrix $\TK$ with $m \times n$ blocks and every block is a convolution matrix, induces several properties, including
\begin{itemize}
\item the number of convolutional kernels in $\TK$ is given by $mn$, and the number of convolutions plus transposed convolutions is $2mn$ per layer. 
\item the block rank of $\TK^\top \TK$ is at most $m$.
\item the number of unique kernels in $\TK^\top \TK$ is at most $(n^2 + n)/2$.
\end{itemize}

Another key observation is that a non-square $\TK$ does not affect the invertibility of the network because the rank-deficient matrix $\TK^\top \TK$ does not need to be inverted for the reverse propagation as defined in \eqref{rev_prop}.

Including the symmetric block-low-rank layer extends the applicability of fully invertible hyperbolic networks to larger input data and an increased number of coarsening stages in the network. See Figure \ref{fig:memorycurves} for a summary of the memory savings.

\subsection{Invertible hyperbolic networks with a different number of input and output channels.}

To segment data into $n_\text{class}$ classes, one typically employs a neural network that maps the data to the last network state (output) $\TY \in \mathbb{R}^{n_x \times n_y \times n_z \times n_\text{chan}}$, where we would choose $n_\text{class} = n_\text{chan}$. The standard accompanying loss function is the multi-class cross-entropy loss
\begin{equation}\label{standard_ce}
L(\TY,\TC) = - \sum_{(i,j,k)}  \sum_{l=1}^{n_\text{chan}} \TC_{i,j,k,l} \log (\TY_{i,j,k,l}),
\end{equation}
where $\TC \in \mathbb{R}^{n_x \times n_y \times n_z \times n_\text{class}}$ represent the labels, and $i$, $j$, $k$ are spatial indices.

In the case of invertible neural networks, the situation is more complicated. The standard form of the fully invertible hyperbolic network outputs the same tensor size as the input, i.e., the number of input and output channels is equal. This is limiting because the data rarely has the same number of input channels as output classes. Fortunately, invertible networks can still be used with a minor modification to the loss function.

Consider data with $n_\text{chan}$ channels input, so there are also $n_\text{chan}$ output channels, assuming the network contains as many forward as inverse transformations. To train a network for segmenting the data into $n_\text{class}$ classes, we need $n_\text{chan} \geq n_\text{class}$. To deal with the mismatch in channel count and the number of classes, we can compute the partial cross-entropy loss

\begin{equation}\label{partial_channel_ce}
L(\TY,\TC) = - \sum_{(i,j,k)}  \sum_{l=1}^{n_\text{class}} \TC_{i,j,k,l} \log (\TY_{i,j,k,l}).
\end{equation}
This loss function simply computes the loss via summation over the $n_\text{class}$ `active' channels only that correspond to labels. The other channels also have an output which is not used to compute a loss or gradient. These channels are still required to reverse the direction of the network, i.e., all channels are required to generate input data from a prediction (and the second to last layer). 

The above assumed that $n_\text{chan} \geq n_\text{class}$, which is generally not the case. The number of input channels can be increased by
\begin{itemize}
\item duplicating (some of) the channels of the input data. This is a valid strategy, but it increases the total memory load as each layer contains more channels. It follows that the input data size grows linearly with the number of duplicated input data channels. 
\item transforming the data into more channels while preserving the information. We can achieve these goals by taking a Haar transform (or another orthogonal transform) that reduces the data resolution and increases the number of channels while keeping the memory load constant. This implies applying the transform to data $\TY$ changes the sizes according to $\TW \TY : \mathbb{R}^{n_x \times n_y \times n_z \times n_\text{chan}} \rightarrow \mathbb{R}^{n_x/2 \times n_y/2 \times n_z/2 \times 8n_\text{chan}}$. Multiple transforms applied in sequence further increase the number of channels if required. To emphasize, this transform is applied to the data before it enters the network. It is, therefore, a one-time operation and does not add significant computational training time. This option is more desirable in terms of memory. 
\end{itemize}

Suppose one is willing to give up invertibility. In that case, we can simply add a non-square linear operator to map the output of the invertible network to the desired number of output channels for the loss, albeit at a higher memory cost and loss of stability guarantees as in Thm. \ref{thm1}.

\subsection{Tasks with different input and output dimensions.}
Applications like hyperspectral land-use segmentation intrinsically reduce the dimensionality between input data and output by collapsing the frequency axis into a point, i.e., $\mathbb{R}^{n_x \times n_y \times n_\text{freq}} \rightarrow \mathbb{R}^{n_x \times n_y}$. In the case of time-lapse hyperspectral segmentation, there is also a reduction along the time axis as $\mathbb{R}^{n_x \times n_y \times n_\text{freq} \times n_t} \rightarrow \mathbb{R}^{n_x \times n_y}$.

The above tasks seem incompatible with the fully invertible hyperbolic neural network that outputs a tensor the same size as the input. This time, we propose to measure the loss over a single slice in the output tensor. That is, embed the known ground-truth labels that depend on the class and spatial coordinates $x$ and $y$ in a larger label tensor $\TC \in \mathbb{R}^{n_x \times n_y \times n_\text{freq} \times n_\text{chan}}$ at slice $p$ as $\TC_{:,:,p,1:n_\text{class}}$. All other entries in the label tensor do not exist and do not contribute to the loss or its gradient computation. See \citet{doi:10.1190/INT-2018-0225.1,doi:10.1190/tle38070534.1} for more information and applications of partial loss functions. The resulting multi-class cross-entropy function reads
\begin{equation}\label{partial_ce}
L(\TY,\TC) = - \sum_{(i,j)}  \sum_{l=1}^{n_\text{class}} \TC_{i,j,p,l} \log (\TY_{i,j,p,l}),
\end{equation}
where $p$ is the fixed tensor-slice index that contains the labels. When only sparse spatial location indices of known labels are available, the sum over $(i,j)$ reduces to a sum over the subset of labeled pixels.

\subsection{Resolution changes between input and output.}
Some applications desire a different resolution for output compared to the input. Once more, straightforward application of the fully invertible hyperbolic network cannot accomplish resolution changes between input and output because the input and output tensors have the same size. It turns out that there is still a way to train an invertible network such that input and output are on different resolutions. To start, consider a single network layer of an invertible hyperbolic network that decreases the resolution via
\begin{equation}
\TY_{3} = 2 \TW \TY_{2} -  \TW \TY_{1} -  h^2 \TK_3^\top \sigma( \TK_3 \TW \TY_{2}), \nonumber
\end{equation}
where $\TW$ is an orthogonal transform that transforms and reorganizes the 4D input as $\TW \TY : \mathbb{R}^{n_1 \times n_2 \times n_3 \times n_\text{chan}} \rightarrow \mathbb{R}^{n_1/2 \times n_2/2 \times n_3/2 \times 8n_\text{chan}}$. Note that the first output channel of the single-level Haar transform is the input on a resolution reduced by a factor of two. Similarly, for other transforms like the pixel-shuffle transform. So as long as the network contains more forward transforms than inverse transforms, the output resolution is lowered by increments of two. Network training can utilize this concept by defining the loss function over a particular selection of output channels. Similar logic applies to resolution increases.

\section{Examples}
The foundation of the following experiments is the fully invertible hyperbolic network, implemented by \cite{philipp_witte_2020_4298853,orozco2023invertiblenetworks}. Its specialization to geoscientific and remote sensing problems is available at \url{https://github.com/PetersBas/FHN_Examples}. The experiments illustrate that the proposed techniques enable the application of fully invertible hyperbolic networks to various large-scale geoscience problems while obtaining satisfactory results. Table \ref{memory} summarizes the memory requirements for the states and convolutional kernels, not including memory allocations for intermediate computations, Lagrangian multipliers for gradient computations and the gradient itself. The table shows that both invertibility and block-low-rank layers are required to run examples on a standard GPU, like a $24$GB \emph{NVIDIA GeForce RTX 3090}. Figure \ref{fig:memorycurves} shows the general memory scaling of our approach. In order to train a non-invertible network with the same number of channels per layer and the same input data size, the number of layers needs to be reduced to keep the memory footprint manageable.

\subsection{Time-lapse hyperspectral land-use change detection}

\begin{figure}[!t]
 	\centering
 	\begin{subfigure}[b]{0.48\textwidth}
 		\includegraphics[width=0.9\textwidth]{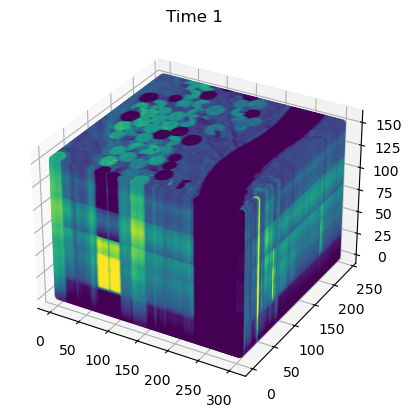}
 		\caption{}
 		\label{fig:Figure1a}
 	\end{subfigure}
 	\begin{subfigure}[b]{0.48\textwidth}
 		\includegraphics[width=0.9\textwidth]{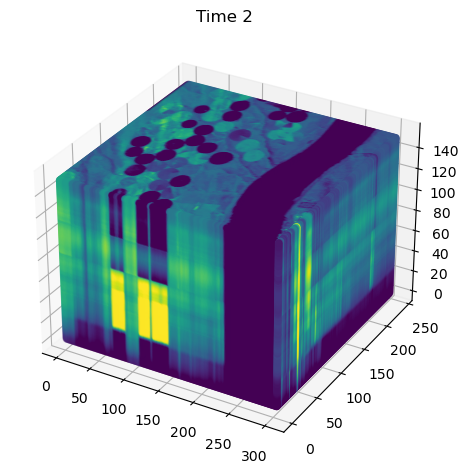}
 		\caption{}
 		\label{fig:Figure1b}
 	\end{subfigure}
 	\caption{Hyperspectral data collected at two different times.}
\label{fig:data}
 \end{figure}

\begin{figure}[!t]
 	\centering
 	\begin{subfigure}[b]{0.3\textwidth}
 		\includegraphics[width=\textwidth]{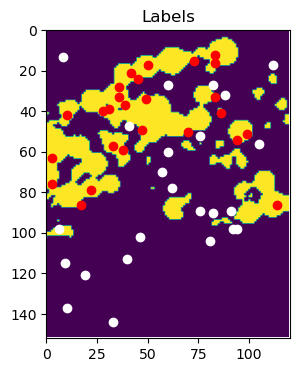}
 		\caption{}
 		\label{fig:Figure2a}
 	\end{subfigure}
 	\begin{subfigure}[b]{0.3\textwidth}
 		\includegraphics[width=\textwidth]{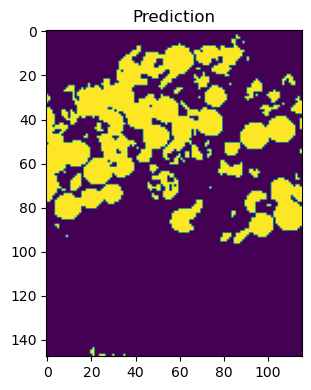}
 		\caption{}
 		\label{fig:Figure2b}
 	\end{subfigure}
 	 	\begin{subfigure}[b]{0.3\textwidth}
 		\includegraphics[width=\textwidth]{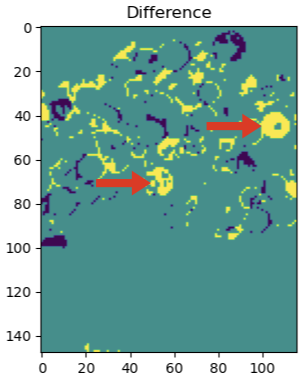}
 		\caption{}
 		\label{fig:Figure2c}
 	\end{subfigure}
 	 	\caption{(a) Plan view of all true labels with point-annotation locations for training and validation overlaid. (b) prediction, and (c) error map. Most errors are boundary effects, and just a few farm fields are identified as changed/not-changed incorrectly (red arrows highlight two examples).}
\label{fig:hyperfigs}
 \end{figure}

The data $\TX \in \mathbb{R}^{n_x \times n_y \times n_\text{freq} \times n_t}$, has two spatial coordinates $n_x$ and $n_y$, the third dimension corresponds to frequency, and there is one channel per time of data collection (two in this example). Figure \ref{fig:data} displays this data set \citep{doi:10.1080/01431161.2018.1466079}. We follow common practice in hyperspectral imaging literature, where part of the segmentation is assumed known. The red and white dots in Figure \ref{fig:hyperfigs} show the $70$ known label locations, and training utilizes $50$ annotations, while the validation is based on the remaining $20$.

\begin{figure}
    \centering
    \includegraphics[width=0.6\linewidth]{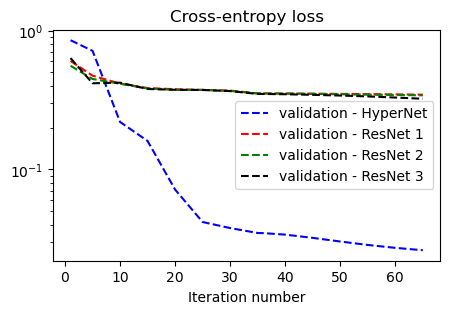}
    \caption{Validation losses for the hyperspectral example, for the proposed network and three different multi-level ResNets.}
    \label{fig:hyperspectral_losses}
\end{figure}
 
This example aims to predict the land-use change on a coarser grid. Table \ref{network_design} contains the network details that correspond to a network that contains one more Haar transform than inverse Haar transforms, so the output feature resolution decreases by a factor of two. We use stochastic gradient descent with momentum and a decaying learning rate for $70$ iterations to minimize the loss function \eqref{partial_ce}. The loss is measured over a few slices of the output tensor that embeds the labels. Figure \ref{fig:hyperfigs} shows true land-use change, prediction, and errors. Aside from some boundary artifacts, a few farm fields were classified incorrectly. The low-memory nature of the invertible network enabled us to input the entire 4D data in one chunk; see also Table \ref{memory} for details regarding memory usage. We also show the validation loss in Figure \ref{fig:hyperspectral_losses}, as well as the validation losses for three comparison multi-level ResNets (see the Appendix for the network details). Because of the memory constraints, the ResNets need to be significantly shorter than the invertible network with BLR layers. The losses for the various ResNets are comparable but do not approach the loss of the proposed network.
 
\subsection{Regional-scale aquifer mapping}

\begin{figure*}[!htb]
\begin{center}
   \includegraphics[width=0.7\textwidth]{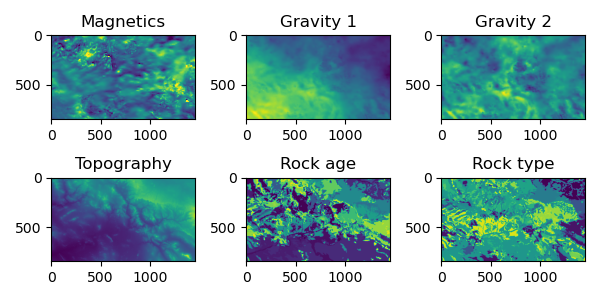}
   \caption{The data inputs for the aquifer mapping example. Each type is placed in a separate channel of the input. We do not use the two geological maps as images. Instead, each class is converted to a map with zero/one values, resulting in 52 separate geological maps. }
   \label{fig:aq_data}
   \end{center}
\end{figure*}

\begin{figure}[!tb]
 	\centering
 	\begin{subfigure}[b]{0.48\textwidth}
 		\includegraphics[width=0.8\textwidth]{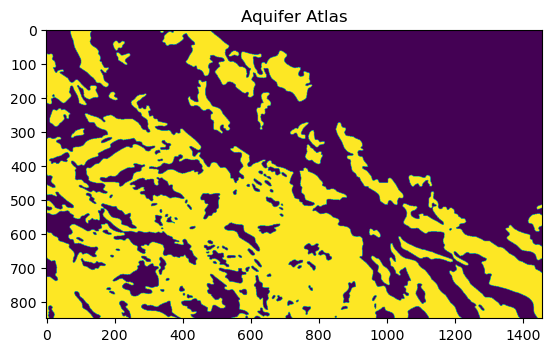}
 		\label{fig:aqr1}
 	\end{subfigure}
 	\begin{subfigure}[b]{0.48\textwidth}
 		\includegraphics[width=0.8\textwidth]{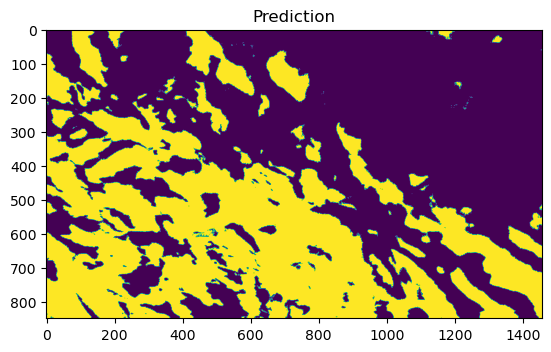}
 		\label{fig:aqr2}
 	\end{subfigure}
 	\begin{subfigure}[b]{0.48\textwidth}
 		\includegraphics[width=0.8\textwidth]{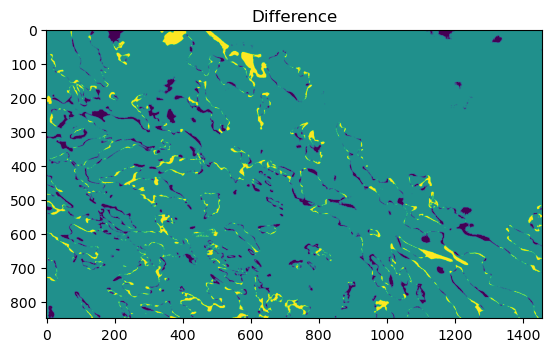}
 		\label{fig:aqr3}
 	\end{subfigure}
 	\caption{True aquifer map, prediction, and difference.}
\label{fig:aq_results}
 \end{figure}

The task is to delineate large aquifers in Arizona, USA; see Figure \ref{fig:aq_results}. The two classes are 1) basin and range / Colorado Plateau aquifer; 2) no aquifer \citep{wateratlas}. The survey area is most of the state. Aircraft and satellite-based sensors collected magnetic data, two types of gravity measurements, and the topography, see Figure \ref{fig:aq_data}. Besides these remotely acquired data, we supplement two types of geological maps: one map in terms of the rock age and one in terms of rock types. The advantage of using geological maps is that they incorporate expert knowledge into our data. Geologists construct these maps by synthesizing their geological knowledge with ground truth observations, hyperspectral data, and various airborne and land-based geophysical surveys. Because the geological maps in Figure \ref{fig:aq_data} are not invariant under the permutation of the class numbers, we use its one-hot encoding (52 separate maps). 

\begin{table}
\begin{center}
\caption{Network designs for the fully invertible networks.}
\label{network_design}
\begin{tabular}{ l c c c}
\multicolumn{1}{c}{} &\multicolumn{1}{c}{} &\multicolumn{1}{c}{\bf Hyperspectral} &\multicolumn{1}{c}{}\\
\multicolumn{1}{c}{\bf Layer}  &\multicolumn{1}{c}{\bf Channels} &\multicolumn{1}{c}{\bf Block rank} &\multicolumn{1}{c}{\bf Feature size}
\\ \hline \\
1-6                & 16 & 16 & $368 \times 288 \times 184$\\
7-18             & 128 & 16 &$184 \times 144 \times 92$
\\ \hline \\
\multicolumn{1}{c}{} &\multicolumn{1}{c}{} &\multicolumn{1}{c}{\bf Aquifer mapping} &\multicolumn{1}{c}{}\\
\multicolumn{1}{c}{\bf Layer}  &\multicolumn{1}{c}{\bf Channels} &\multicolumn{1}{c}{\bf Block rank} &\multicolumn{1}{c}{\bf Feature size}
\\ \hline \\
1-4               & 112 & 24 & $848 \times 1456$\\
5-7               & 448 & 24& $424 \times 728 $\\
8-10              & 1792 & 24& $212 \times 364 $\\
11-28             & 7168 & 24& $106 \times 182 $\\
29-32             & 1792 & 24& $212 \times 364 $\\
32-34             & 448 & 24& $424 \times 728 $\\
35-39             & 112 & 24& $848 \times 1456 $
\\ \hline \\
\end{tabular}
\end{center}
\end{table}

\begin{figure}[!t]
 	\centering
 	\begin{subfigure}[b]{0.48\textwidth}
 		\includegraphics[width=0.8\textwidth]{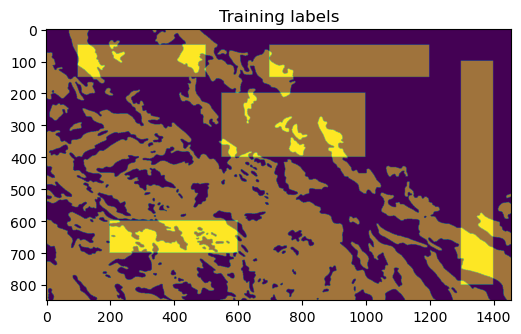}
 		\label{fig:aql1}
 	\end{subfigure}
 	\begin{subfigure}[b]{0.48\textwidth}
 		\includegraphics[width=0.8\textwidth]{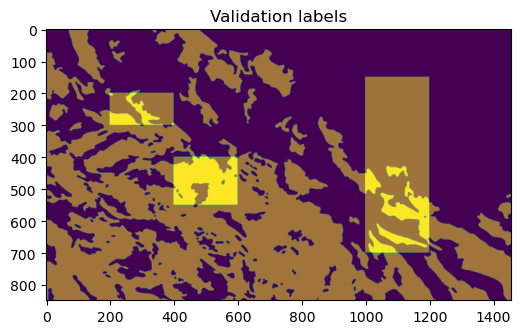}
 		\label{fig:aql2}
 	\end{subfigure}
 	\caption{Locations of the training and validation labels for the aquifer mapping example.}
\label{fig:aq_labels}
 \end{figure}
 
The experimental setting assumes an expert annotated the aquifers in a few patches, see Figure \ref{fig:aq_labels}, so the neural network assists a domain expert by interpolating and extrapolating limited annotation. Training is similar to the previous example: SGD with momentum with a decaying learning rate for $140$ iterations to minimize the multi-class cross-entropy loss. Each iteration uses about $10\%$ of the known labels (randomly selected) to compute an approximation of the loss and the gradient. We also augment the data with random flips and permutations. The network details can be found in Table \ref{network_design}. The $39$ layer fully invertible hyperbolic network uses three coarsening stages (see table \ref{network_design}) to increase the receptive field size and enable information to propagate over larger spatial distances. Figure \ref{fig:aq_results} displays the results and errors. Most of the errors are concentrated along some of the geological rock-type boundaries. 

This example showed that neural networks could assist domain experts and mimic their work for aquifer mapping. Invertible networks can deal with such large computational domains with many input channels in one chunk. The invertibility reduced the memory required for storing the network states from $21.02$GB to just $1.66$GB, see table \ref{memory}. For this example with many input channels and multiple coarsening stages, training the parameters of the network in a compressed/factorized form directly used just $0.12$GB for storing the convolutional kernels instead of the $32.19$GB for storing unreasonably many convolutional kernels that a standard invertible hyperbolic network would require for this particular design, see table \ref{memory}.

\subsection{3D interpolation-segmentation of a seismic image volume from borehole information.}
Building a 3D geological model from seismic imaging means grouping several layers, structures, or geological units to obtain a simplified geological model that conveys the information of interest. The interpretation of seismic volumes is challenging due to imaging artifacts (data noise, violation of assumptions in the imaging algorithm, poor illumination from seismic waves), spatial variation in the appearance of the interfaces, discontinuities of the interface due to geological faults, and a lack of ground-truth away from boreholes. 

\begin{figure}[!htb]
 	\centering
 	\begin{subfigure}[b]{0.45\textwidth}
 		\includegraphics[width=\textwidth]{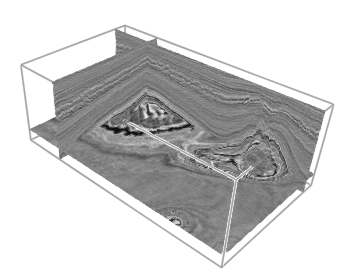}
 		\caption{}
 		\label{fig:seis_data3d}
 	\end{subfigure}
 	\begin{subfigure}[b]{0.45\textwidth}
 		\includegraphics[width=\textwidth]{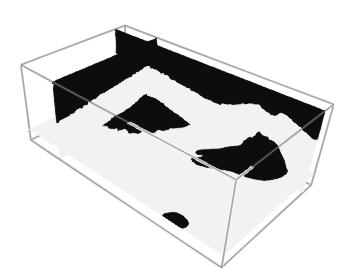}
 		\caption{}
 		\label{fig:seis_labels3d2}
 	\end{subfigure}\\
    \begin{subfigure}[b]{0.45\textwidth}
        \includegraphics[width=\textwidth]{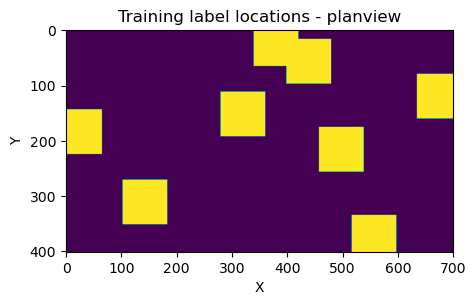}
        \caption{}
        \label{fig:seis_planview_train}
    \end{subfigure}
    \begin{subfigure}[b]{0.45\textwidth}
        \includegraphics[width=\textwidth]{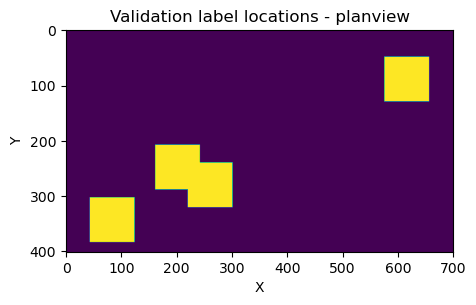}
        \caption{}
        \label{fig:seis_planview_val}
    \end{subfigure}
 	 	\caption{(a) Full 3D data; training uses sub-cubes. (b) fully labeled data volume, training and validation use small parts of the labels, as indicated by the plan-view figure (c) of the computational domain and highlighted are the training/validation locations (small areas near borehole locations). The remainder of the labels are used for testing purposes only.}
\label{fig:seis_labels}
 \end{figure}

Here, we present an experiment aiming to segment the full 3D seismic volume (Figure \ref{fig:seis_labels}) from borehole information. We assume that a small area around each borehole was interpreted by an expert and can serve as training and validation labels (Figure \ref{fig:seis_labels}). Interpreting the seismic image close to boreholes is relatively easy due to the proximity to the ground truth.

While seismic interpretation from borehole information is nothing new, most work maps 2D image-to-image. Some 3D seismic interpretation approaches operate 3D-to-3D but are limited to training with relatively small 3D sub-cubes due to memory limitations and use sub-cubes of up to $128^3$ \citep{doi:10.1190/geo2018-0646.1,doi:10.1190/segam2019-3216307.1,doi:10.1190/segam2018-2997304.1,doi:10.1190/INT-2018-0224.1,10.1190/geo2020-0945.1,9851460}. Using larger 3D sub-cubes enables learning larger-scale structures present in the data. \citet{peters2020fully} show the first 3D interpretation approach using an invertible network and use an input size of up to $192 \times 192 \times 288 \times 3$. Here, we show results from training on the largest inputs to date that are about $5.7\times$ larger than previous work.

\begin{figure*}[!t]
\begin{center}
   \includegraphics[width=0.99\textwidth]{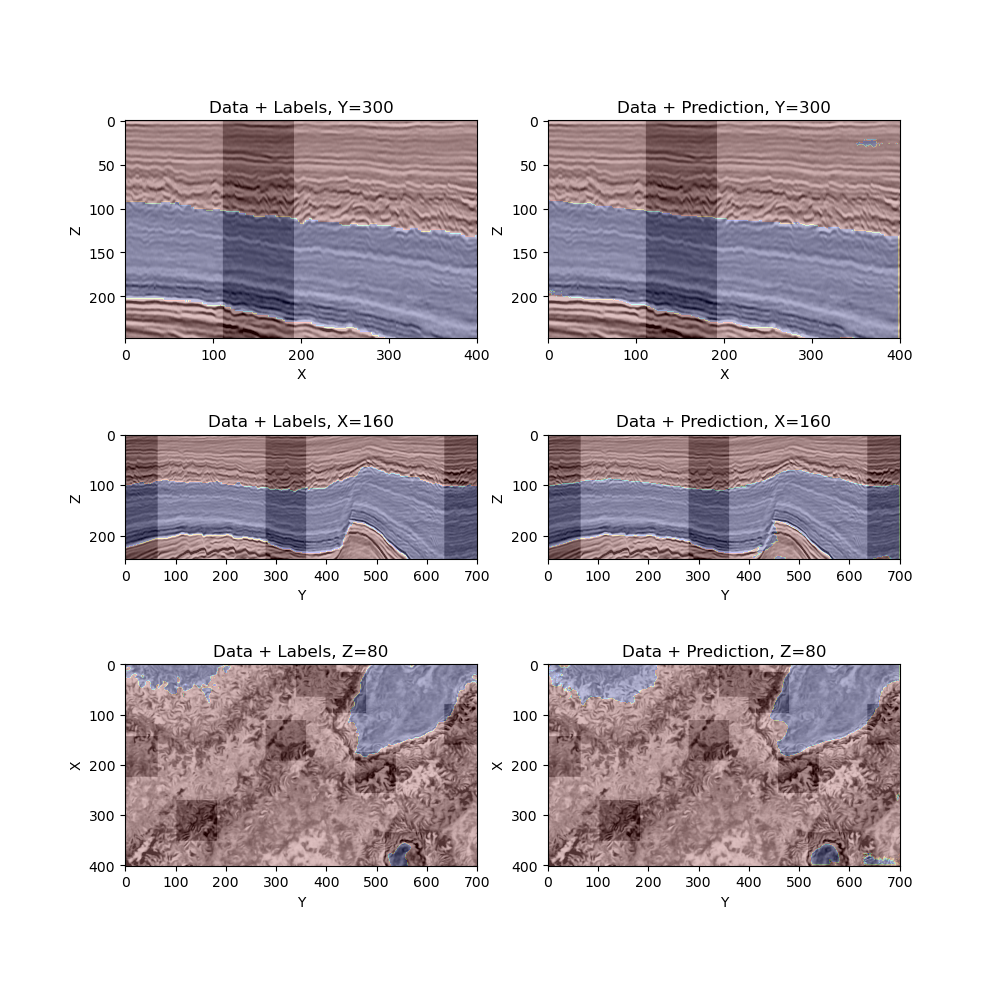}
   \caption{Three orthogonal cross-sections from the final prediction and the true labels. Training label locations are shaded.}
   \label{fig:seis_results}
   \end{center}
\end{figure*}

The full data has a size of $401 \times 701 \times 248$. For training, we replicate the data into $12$ channels and input to the network a randomly selected sub-cube of size $248 \times 248 \times 248 \times 12$. Table \ref{network_design_seismic} lists the network details. We reduce the cross-entropy loss using the ADAM optimizer for $240$ iterations with a decaying stepsize. Each iteration selects a randomly located sub-cube. Table \ref{seismic_IoUs} displays the final results' intersection over union (IoU). Figure \ref{fig:seis_results} shows 2D cross-sections from the 3D volume, accompanied by the true labels. The final prediction results from inference on the full data volume split into a couple of overlapping pieces. This example shows that we can obtain good segmentation results from partial labeling while training on large 3D input sub-cubes. The size of the sub-cubes is important because selecting only small sub-cubes would lead to most cubes containing no labels. Larger sub-cubes can connect more of the data to labeled locations.

\begin{table}[!tb]
\caption{Network design for the fully invertible network for the 3D seismic interpretation example.}
\label{network_design_seismic}
\begin{center}
\begin{tabular}{lclclcll}
\multicolumn{1}{c}{\bf Layer}  &\multicolumn{1}{c}{\bf Block rank} &\multicolumn{1}{c}{\bf Channels} &\multicolumn{1}{c}{\bf Feature size} \\
\hline \\
1-2     & 8     &   12         & $248 \times 248 \times 248$ \\
3-5     & 16   &   96      & $124 \times 124 \times 124$ \\
6-8     & 32   &  768     & $62 \times 62 \times 62$ \\
9-18   & 32  &  6144   & $31 \times 31 \times 31$ \\
19-21 & 32 &  768      & $62 \times 62 \times 62$\\
22-24 & 16 &   96       & $124 \times 124 \times 124$ \\
25-30 & 8   &   12          & $248 \times 248 \times 248$ \\
\end{tabular}
\end{center}
\end{table}

Table \ref{memory} displays the memory usage of our network and equivalent standard invertible and non-invertible networks. The table also shows that a direct comparison with the equivalent non-invertible version of the hyperbolic network, and without block-low-rank layers is not possible on a standard $\approx24$ GB GPU. Instead, we evaluated two closely related networks that just fit on the GPU. These networks are similar to ours (Table \ref{network_design_seismic}), except with one less level and a few layers shorter. See the Appendix for network details and Figure \ref{fig:seis_results_compare} for prediction images. Table \ref{seismic_IoUs} contains the statistics in terms of intersection over union (IoU), which shows that using one level less or reducing the number of layers to make the network fit/trainable on a GPU, comes at the cost of a significant drop in IoU score.

\begin{table}[!tb]
\caption{Validation IoU values for the 3D seismic example. Invertible + BLR layer network design from Table \ref{network_design_seismic} is according to Equation \eqref{network}. The memory requirements for this network and its non-invertible equivalent without block-low-rank layers are shown in Table \ref{memory}. See the Appendix for the design details of networks $^a \& ^b$.}
\label{seismic_IoUs}
\begin{center}
\begin{tabular}{ll}
\bf Network type & \bf Validation IoU\\
\hline 
Invertible + BLR layers (ours) & class 1/class 2:  $0.97$/$0.96$\\
Non-invertible equivalent, no BLR layers & out of memory \\
Largest non-invertible related$^a$ & class 1/class 2: $0.91$ / $0.85$ \\
Largest non-invertible related$^b$ & class 1/class 2: $0.92$ / $0.88$ \\
\end{tabular}
\end{center}
\end{table}

This example also showed that multi-level, fully invertible hyperbolic networks are suitable for 3D seismic segmentation and that the more than six thousand channels are no issue if we use symmetric layers with a low block-rank.

\subsection{Effect of the selection of the maximum block-rank}
The previous examples utilize a block rank of $\TK^\top \TK$ that is much below full rank. Naturally, questions arise about selecting the block rank and determining the sensitivity of the network performance to the block rank. 

\begin{figure}[!htb]
 	\centering
 	\begin{subfigure}[b]{0.25\textwidth}
 		\includegraphics[width=\textwidth]{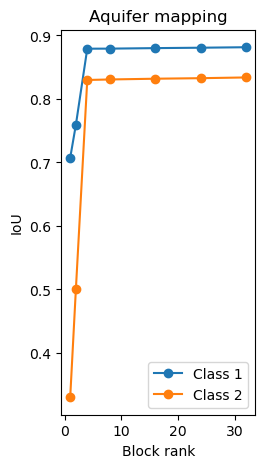}
 		\label{fig:AQ_BLR}
 	\end{subfigure}
 	\begin{subfigure}[b]{0.25\textwidth}
 		\includegraphics[width=\textwidth]{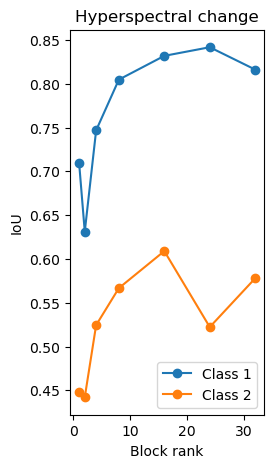}
 		\label{fig:HYP_BLR}
 	\end{subfigure}
    \begin{subfigure}[b]{0.25\textwidth}
        \includegraphics[width=\textwidth]{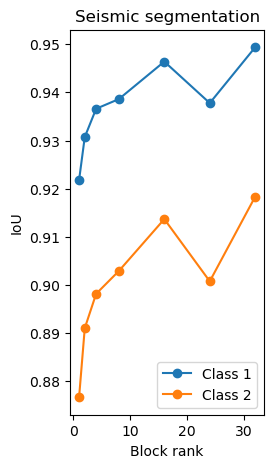}
        \label{fig:seis_BLR}
    \end{subfigure}
 	 	\caption{Plots of the IoU versus the selected block-rank.}
\label{fig:IoU_vs_BLR}
 \end{figure}

The maximum rank we can practically select is limited by the available memory to store convolutional kernels, or by the computational time that is available. Figure \ref{fig:IoU_vs_BLR} provides a more quantitative and experimental answer. The figure shows all experiments repeated for various values of the maximum block-rank and averaged over five random initializations of the network parameters. The conclusion is that a very low block rank generally comes at the cost of slightly reduced prediction quality, as measured using intersection over union. A high block rank can also slightly reduce the prediction quality because it generates less implicit regularization (no other forms of regularization were used in the numerical experiments.) We finalize the experimental evaluation by noting that the IoU results show clear trends, but the assessment of the geoscientific and remote sensing examples remains challenging because the seismic, aquifer, and hyperspectral labels are expert interpretations of the data, and not ground truth. The aquifer mapping example is based on various data sources, including ones not fed into the network. All of these come with some ambiguity.

\section{Conclusions}
The high memory requirements for training a deep neural network using automatic differentiation is a critical issue that limits the application to large input-data blocks like hyperspectral data, very large-scale multi-modality 2D geoscientific maps and airborne-remote sensing, as well as 3D seismic imagery. Fully invertible neural networks mostly solve the memory requirement issues for network states and achieve a constant memory footprint independent of the number of layers and pooling/coarsening stages inside the network.

This work takes a closer look at the fully invertible hyperbolic network based on a conservative leapfrog discretization of the non-linear hyperbolic telegraph equation. Problems are uncovered that prevent the direct application of fully invertible hyperbolic networks to tasks that require multiple coarsening/pooling stages in the network, applications that reduce a 3D/4D tensor into a 2D map of the earth, tasks with different numbers of input and output channels, and applications with resolution changes. For each of these issues, we provided a solution that enables the application of the network without fundamentally altering its design. 

We introduce a layer design where the matrix representation of the convolutional kernels has a low block-rank structure. This design changes the exponential growth of the memory for convolutional kernels as a function of the number of pooling stages into a tuneable memory footprint. Changing the number of channels, dimensions, and resolution between input and output is enabled by embedding the labels into larger tensors in combination with particular ways to measure the loss, and using a different number of forward and inverse orthogonal transforms inside the network.

Examples illustrate how to apply fully invertible networks to time-lapse hyperspectral data with a resolution change, very large-scale multi-modal remote sensing for sub-surface aquifer mapping, and 3D seismic interpretation. The tools developed in this work enable invertible networks to be applied to these problems and thus learn from larger input blocks of data, which in turn enables the network to learn from larger-scale structures present in the data.

\clearpage

\bibliographystyle{unsrtnat}
\bibliography{biblio,VideoRefs,HyperSpectralRefs,seismic_horizon}

\appendix
\section{Additional experimental details}
Here, we provide more details regarding the comparison network designs. Table \ref{memory} shows the memory requirements for the fully invertible hyperbolic network with block-low-rank (BLR) layers. Table \ref{memory} also shows the memory in case we do not use the invertibility of the network to compute gradients, and if we do not use BLR layers. Details for the comparison networks in the numerical experiments section are provided below.

\subsection{Hyperspectral time-lapse}
We compared the results of our proposed network to the symmetric ResNet \citep{HaberRuthotto2017a}
\begin{equation}\label{sym_resnet}
\TY_{j} = \: \TY_{j-1} - h \TK(\Ttheta_{j})^\top \sigma( \TK(\Ttheta_{j}) \TY_{j-1}).
\end{equation}
The hyperspectral example requires a two-level network with one pooling operation. We employ the BLR layers, as in the fully invertible hyperbolic network. However, because the ResNet relies on automatic differentiation and not on invertibility for gradient computations, it is impossible to fit the same number of network layers on the GPU. Therefore, we have to shorten the network. Figure \ref{fig:hyperspectral_losses} compared three ResNets, and Table \ref{tab:HyperSpectralResNets} describes the design details.

\begin{table}[htb]
    \centering
    \begin{tabular}{cccc}
    \multicolumn{4}{c}{\bf Comparison ResNets for the Hyperspectral example}\\
    \hline\\
        Network & $\#$ layers level 1 & $\#$ layers level 2 & \\
        1 & 3 & 4 & Fig. \ref{fig:hyperspectral_losses}\\
        2 & 2 & 5 & Fig. \ref{fig:hyperspectral_losses}\\
        3 & 5 & 2 & Fig. \ref{fig:hyperspectral_losses}\\
        4 & 6 & 1 & out-of-memory\\
        5 & 3 & 5 & out-of-memory\\
        6 & 5 & 3 & out-of-memory\\
    \end{tabular}
    \caption{Network designs for the comparison ResNets designs according to Eq. \ref{sym_resnet}. These are all two-level networks, i.e., a few ResNet blocks followed by a pooling operation and some more ResNet blocks.}
    \label{tab:HyperSpectralResNets}
\end{table}

\subsection{3D Seismic Interpretation}
\begin{figure}[htb]
\begin{centering}
\begin{subfigure}[htb]{0.65\textwidth}
   \includegraphics[width=0.99\textwidth]{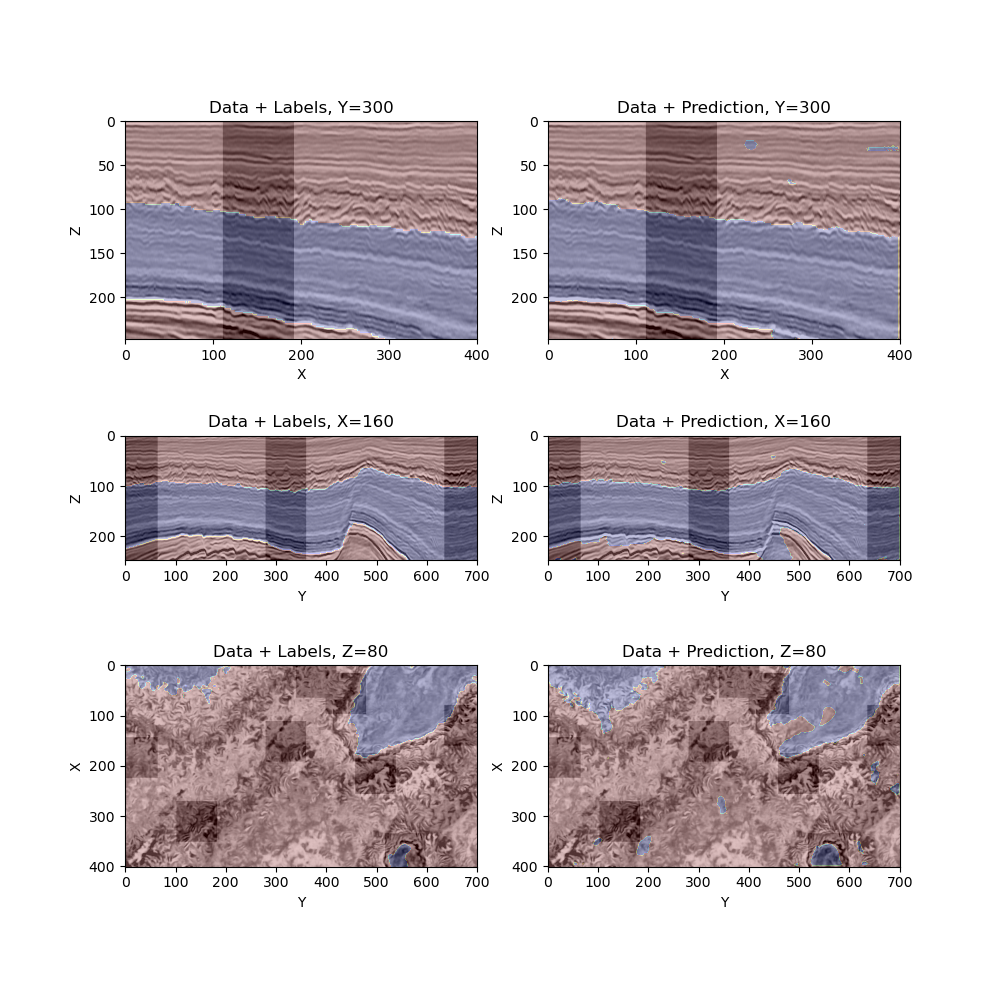}
   \end{subfigure}
\begin{subfigure}[htb]{0.65\textwidth}
   \includegraphics[width=0.99\textwidth]{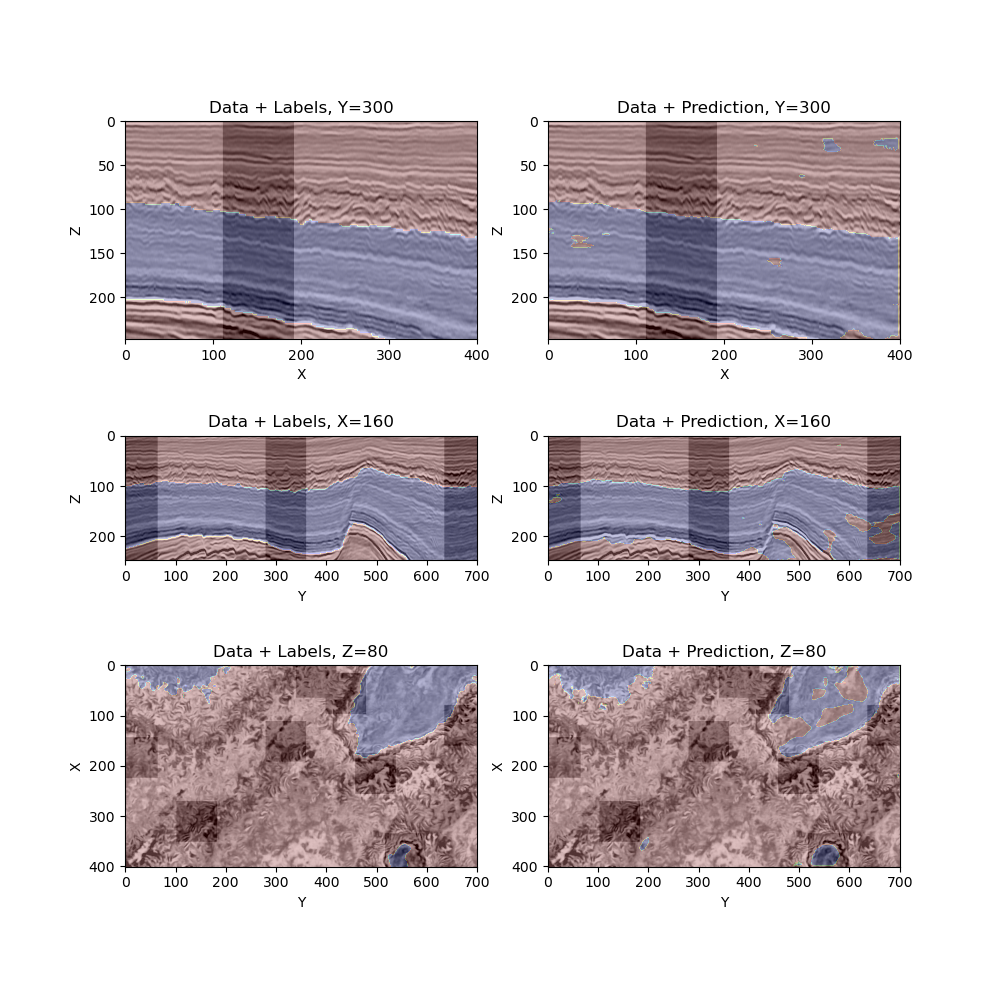}
   \end{subfigure}
   \caption{Predictions using the comparison networks form Table \ref{network_design_seismic_comparables}, network \emph{a} (top) and network \emph{b} (bottom). Figures show three orthogonal cross-sections from the final prediction and the true labels. Training label locations are shaded. Both results look mostly realistic, but contain more holes and incorrectly assigned patches compared to Figure \ref{fig:seis_results}.}
   \label{fig:seis_results_compare}
\end{centering}
\end{figure}

Memory-wise, it is impossible to train the seismic segmentation example on most GPUs if we are not using invertibility and BLR kernels, because the convolutional kernels then require 41.16 GB and just the states require 21.96 GB. These numbers do not include memory for intermediate computations, the gradient itself, and other storage for the optimizer. For a comparison, we construct the largest possible networks that fit on a $24$GB GPU, that are as similar as possible to the network in Table \ref{network_design_seismic}. We modify the design by reducing the number of levels by one level, and shortening the network. Network \emph{a} uses BLR layers, while network \emph{b} does not. We refer back to Table \ref{seismic_IoUs} for the performance metrics, which show the networks with fewer layers and one level less significantly underperform the proposed network. See Figure \ref{fig:seis_results_compare} for predictions from the shorter network.

\begin{table}[!tb]
\caption{Network designs that just fit on a $24$GB GPU, to compare to our fully invertible network for the 3D seismic interpretation example.}
\label{network_design_seismic_comparables}
\begin{center}
\begin{tabular}{lclclcll}
\multicolumn{1}{c}{\bf Layer}  &\multicolumn{1}{c}{\bf Block rank} &\multicolumn{1}{c}{\bf Channels} &\multicolumn{1}{c}{\bf Feature size} \\
\hline \\
\multicolumn{4}{c}{\bf network a}\\
1-2     & 8     &   12         & $248 \times 248 \times 248$ \\
3-5     & 16   &   96      & $124 \times 124 \times 124$ \\
6-17     & 32   &  768     & $62 \times 62 \times 62$ \\
18-20     & 16   &   96      & $124 \times 124 \times 124$ \\
21-26    & 8   &   12          & $248 \times 248 \times 248$ \\
\hline \\
\multicolumn{4}{c}{\bf network b}\\
1-2     & 12     &   12         & $248 \times 248 \times 248$ \\
3-4     & 96   &   96      & $124 \times 124 \times 124$ \\
5-6     & 768   &  768     & $62 \times 62 \times 62$ \\
7-8     & 96   &   96      & $124 \times 124 \times 124$ \\
9-14    & 12   &   12          & $248 \times 248 \times 248$ \\
\hline

\end{tabular}
\end{center}
\end{table}

\end{document}